\documentclass[11pt,twoside]{article}
\usepackage{asp2010}

\resetcounters

\begin{document}

\title{Filtergraph: A Flexible Web Application for Instant Data Visualization of Astronomy Datasets}
\author{Dan Burger$^1$, Keivan G.\ Stassun$^{1,2}$, Joshua A.\ Pepper$^1$,
Robert J.\ Siverd$^1$, Martin A.\ Paegert$^1$, Nathan M.\ De Lee$^1$
\affil{$^1$Department of Physics \& Astronomy, Vanderbilt University, VU Station B 1807, Nashville, TN 37235}
\affil{$^2$Department of Physics, Fisk University, 1000 17th Ave.\ N., Nashville, TN 37208}}

\begin{abstract}
Filtergraph is a web application being developed by the Vanderbilt Initiative
in Data-intensive Astrophysics (VIDA) to flexibly handle a large variety
of astronomy datasets. While current datasets at Vanderbilt are being used
to search for eclipsing binaries and extrasolar planets, this system can
be easily reconfigured for a wide variety of data sources. The user loads
a flat-file dataset into Filtergraph which instantly generates an interactive
data portal that can be easily shared with others. From this portal, the user
can immediately generate scatter plots, histograms, and tables based on the
dataset. Key features of the portal include the ability to filter the data
in real time through user-specified criteria, the ability to select data by
dragging on the screen, and the ability to perform arithmetic operations
on the data in real time. The application is being optimized for speed in
the context of very large datasets: for instance, plot generated from a stellar
database of 3.1 million entries render in less than 2 seconds on a
standard web server platform. This web application has been created using
the Web2py web framework based on the Python programming language. 
Filtergraph is freely available at \url{http://filtergraph.vanderbilt.edu/}.
\end{abstract}

\section{Introduction}
Increasingly in astronomy there is a need for performing quick-look
inspection and visualization of large datasets in order to: easily
ascertain the nature and content of the data, to begin identifying
possibly meaningful structures or patterns in the data, and in order to guide
more computationally costly deep-dive analyses of the data. For example,
consider that the first products of a large survey project is often a
large database with many columns (representing the various measurable
and/or derived quantities) and with many rows (representing the individual
objects of study); it is not uncommon for such databases to include tens
of columns and millions of rows.

To even begin visualizing such datasets---let alone perform basic analyses---can
be a daunting task. The researcher is faced with questions such as: What
is the content and what does it look like? Where are the ``holes" in the
data (missing or bad data) and are there systematics or biases to be wary
of? What are the relationships among the variables in the dataset, and
are they meaningful? Are there interesting patterns that might be worthy
of deeper investigation and analysis?

\includegraphics[width=120mm]{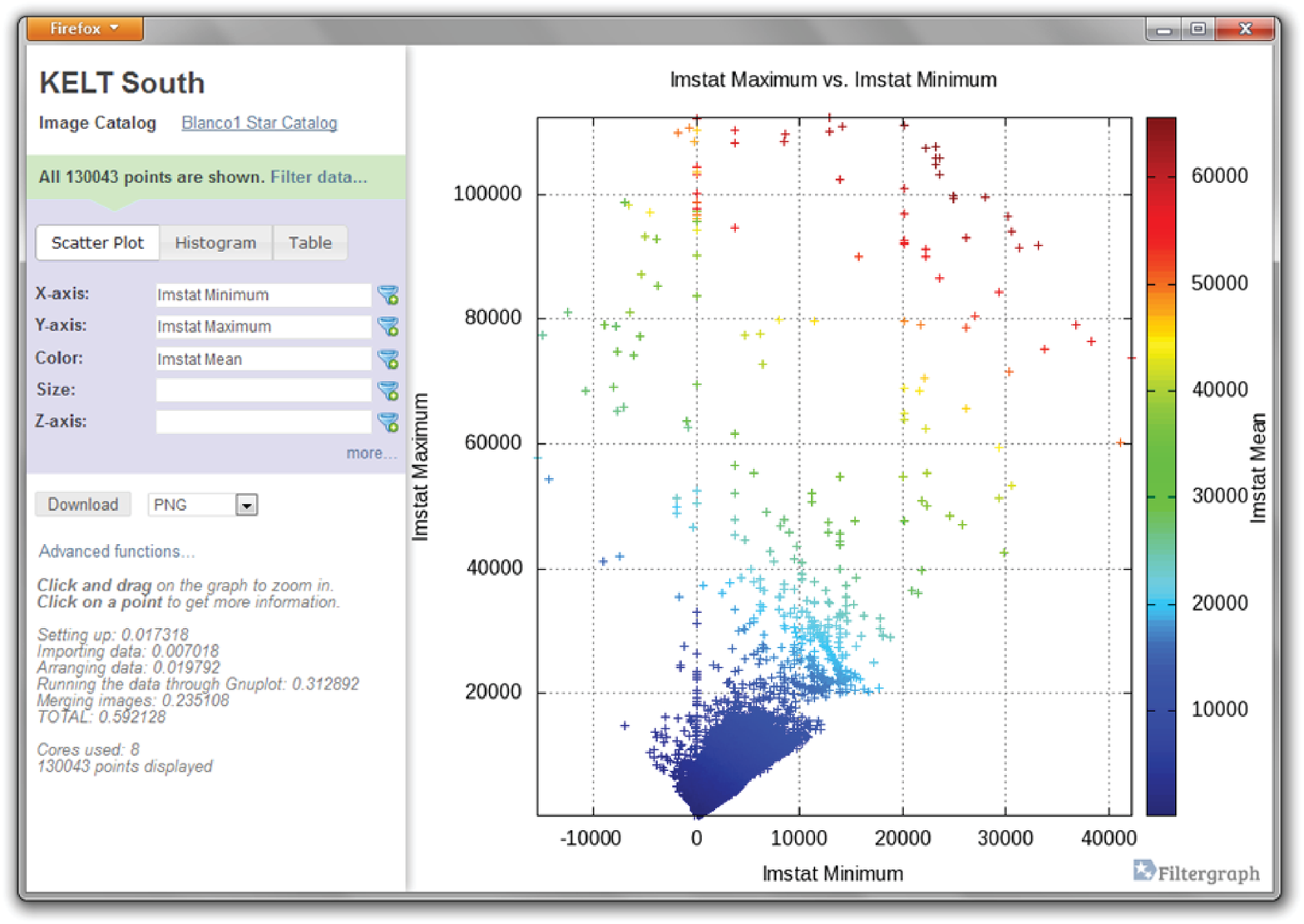}

Indeed, the sheer size of such datasets and their multidimensionality
is at the heart of what makes their visualization so challenging. To
identify potentially meaningful patterns often requires ``seeing" the data
simultaneously across multiple dimensions and with appropriate ``slices"
through multiple multidimensional spaces.

Arguably even more fundamental to the visualization challenge is what might
be called the high ``potential barrier" that the user faces to even begin the
visualization process. Certainly there exist high-end tools for databasing,
plotting data, and so on. But for many researchers there is from the first
instance a very large overhead associated with using such tools: importing
the data and correctly specifying meta-data, keeping track of a large
number of variable names, issuing and scripting commands for plotting, plotting pairwise variables against one another over multiple iterations, attempting
to filter out bad data, re-rendering plots to restricted data ranges,
attempting to represent multiple variables at once, and so on. Faced with
such high overhead in time and effort, there is the temptation to either
skip the crucial quick-look visualization step altogether, or else to make
very limited attempts at representing the data with simple plots based on
preconceptions about what should be meaningful to visualize.

We have developed Filtergraph as an easy-to-use web-based solution to
this problem. The principal motivation for Filtergraph is to make the
``activation energy" of beginning the data visualization process as small
as possible. Users can register to use the tool instantly, can immediately
upload datasets without the requirement of meta-data specification, and
can thus begin seeing their data in seconds. We have also sought to
make Filtergraph intuitive. Plots involving 2, 3, 4, or 5 dimensions (3D +
color + symbol size), as well as histograms, can be generating with a single
click, and variable name fields are pre-filled and auto-complete so that
the user does not need to remember the full content of the dataset in order
to make plots. Mathematical operations on individual variables---or indeed
among variables---can be performed on the fly. (For example, one can simply
specify the x-axis of a plot as {\tt bmag}$-${\tt vmag}.) 
And, true to its name, data
can be easily filtered, for example by specifying data ranges explicitly
or by dragging over a region of interest on the screen, rotating a plot,
and other similarly intuitive gestural commands.

Importantly, Filtergraph is fast. There is nothing more frustrating
or deterrent of the creative visualization process than being faced
with long lag times between subsequent plot renderings. If plots do not
update instantly, users will be more likely to avoid the penalty
associated with trial and error exploration. Filtergraph renders a plot
of 1.5 million points in approximately one second, enabling and encouraging
natural, seamless interaction with and exploration of the data. The user
can change variables, attempt different mathematical operations, filter
in and out, move back and forth between different representational forms,
over and over, in seconds and without cognitive interruption.

Finally, Filtergraph is designed to facilitate sharing. Any Filtergraph plot
can be saved as a graphics file in various formats. Filtered subsets of
the data can also be saved as tables in various formats. More importantly,
each dataset is instantly set up as a sharable ``data portal" ({\tt http://filtergraph.vanderbilt.edu/yourname})
that can be provided to collaborators. Instead of sending collaborators a copy of
the raw data file, the user can easily provide a simple URL that contains the
data and the ability to instantly visualize it, thus greatly facilitating the collaboration process.

\section{Current capabilities and use}
Filtergraph is a free web-based service. By uploading a
dataset to Filtergraph, the user is presented with a portal that can be used to generate
plots and tables based on the dataset. The portal can be shared with others
and provides interactive features such as zooming and obtaining information
about a point on the scatter plot.

Filtergraph currently supports visualization in three forms. A scatter plot consists of at least two columns; one for the X-axis and one for the Y-axis. Three optional axes may be set. The color axis sets the color of
each point based on the color spectrum, with higher values receiving dark
red and lower values receiving dark blue. The size axis sets the size of
each point with higher values receiving larger points and lower values
receiving smaller points. The Z-axis transforms the scatter plot into a
three-dimensional plot which can be viewed from different angles. Additionally, the data can be plotted as a histogram based on one axis, split based
on a set number of bins. Filtergraph can also return the data as a table based on a selected subset of rows and columns
in the dataset. This is particularly useful for target selection.

Currently Filtergraph works with ASCII data files only, but we are expanding
the tool to allow datasets in FITS tables, XML, and other common formats. We
began with ASCII because this file format is a universal standard: files
can be generated on one computer and transferred to another computer easily,
regardless of the operating system used. In addition, ASCII data files can
be read and written using a variety of programming languages, often without
needing to import a special library. A typical astronomy dataset consists of
a header followed by the data itself. The header is one line and consists
of the names of all the columns, each separated by one or more whitespace
characters (spaces and/or tabs). The header may optionally start with the
pound symbol (\#). Since whitespace characters are used to separate the column
names, the column names may not contain whitespace characters themselves.

The rest of the file contains the actual data, with each row separated by a
newline character. The data may consist of strings, integers, floating-point
numbers, and dates. It is not necessary for the dataset to include the
type of information present.

A Filtergraph portal consists of two parts, with the left sidebar being
used to control the main content. Key components of the left sidebar include: the name and description of the portal; the ability to switch between datasets on the portal, if more than one dataset is available; the ability to apply criteria, or "filters", on the dataset; the ability to switch between the scatter plot, histogram, and table modes; the ability to apply display settings; the ability to output the data to a file (PNG, JPEG, GIF, Postscript, and PDF for graphs, HTML and ASCII for table data); and additional instructions and status info for using the dataset.

Axes are changed using a drop-down box that can be edited for advanced functions. The
following advanced functions are supported: addition, subtraction, multiplication, division, modulo, power, natural logarithm, base 10 logarithm, absolute value, square root, exponential, pi, sine, cosine, tangent, and the hyperbolic and inverse versions of sine, cosine, and tangent.

For scatter plots and histograms, the main content can be modified
interactively. Clicking on a point on the graph displays a popup window with all
of the data for that particular point. For some portals, clicking on the
point can grab data from external services such as the SDSS image libraries. Additionally, clicking and dragging on the window allows the user to zoom in on a particular region of the graph. A link is then provided to reset the zoom. Also, when the Z-axis is enabled for scatter plots, the user can rotate
the 3-D scatter plot.

Filtergraph also contains an administration interface for maintaining multiple portals which contain one
or more datasets. An administrator for a portal can change or delete the
portal and add, change or delete datasets associated with the portal. An
administrator can also select another user to co-administer the portal.

Once a user registers for the site, he or she is asked to create a portal. To
create a portal, the user provides a name and URL and then uploads an
initial dataset. This dataset is then inspected to determine the data
types for each column. At this point, the user may provide alternate names
for each header in the dataset. If the dataset is formatted correctly based on provided guidelines,
the portal is then ready for use. Filtergraph also provides many standard user administration features provided by the Web2py web framework, such as changing profile information and obtaining
lost usernames and passwords.

\section{Technical details of Filtergraph}
Filtergraph was written in Python and developed using the Web2py framework. Once the data is loaded and arranged using Numpy, the results are distributed among multiple processes of Gnuplot and then flattened using Graphicsmagick. The number of Gnuplot processes used depends on the amount of data that is being plotted. In addition, JQuery, JQueryUI, and ImgAreaSelect are Javascript libraries that are used to enhance the web browser experience. 

\acknowledgements 
The authors would like to acknowledge the users of Filtergraph who have provided valuable feedback throughout the development of this project, and support 
through a NASA ADAP grant and from the Vanderbilt Initiative in Data-intensive
Astrophysics (VIDA): {\tt www.vanderbilt.edu/astro/vida}. 

%\bibliography{aspauthor}

\end{document}